\documentstyle{article}

\begin{document}
\hfill SMC-PHYS-164

\hfill hep-th/0009091

\vskip 15pt
\centerline{\large\bf	Unified Field Theory Induced on the Brane World }
\vskip 10pt
\centerline{                Keiichi Akama }
\centerline{\small\sl       Department of Physics, Saitama Medical College,
                Kawakado, Moroyama, Saitama, }
\centerline{\small\sl       350-0496, Japan}
\centerline{\small       E-mail: akama@saitama-med.ac.jp}
\vskip 5pt

\leftskip 15mm\rightskip 15mm

We show how the gravity, gauge, and matter fields are induced on dynamically localized brane world.

\vskip15pt
\leftskip 0mm\rightskip 0mm

This talk is based on a work 
	done in collaboration with 
	A.~Akabane, T.~Hattori, K.~Katsuura, and H.~Mukaida.\cite{aahkm} 
Here I would like to show 
how the gravitational, gauge and matter field theories are induced on a brane world.
Recently, idea of the brane world attracts much attention
	with expectation for large extra dimensions, 
	or in connection with the superstring theory.\cite{ADD}$^-$\cite{RS}
The idea that we live in a hyper-surface in a higher dimensional spacetime 
	is very old.\cite{early}$^-$\cite{Akama87}
The present author also presented a model eighteen years ago.\cite{Nara}

Here, we consider a three dimensional domain wall     
localized by the double well potential 
\begin{eqnarray}
U=-\frac{1}{4}\lambda\left(\Phi^2-\frac{m^2}{\lambda}\right)^2,
\end{eqnarray}
for a scalar field $\Phi$ in 4+1 dimensional spacetime,\cite{RubShap}
	where 
	$\lambda$ and $m$ are constants.
The well known kink solution 
\begin{eqnarray}
\Phi=\Phi_K(y)\equiv\frac{m}{\sqrt{\lambda}}\tanh\frac{my}{\sqrt2},
\end{eqnarray}
	for this model 
	gives rise to a flat domain wall located at the $y$=0 hyper-surface.

Such a domain wall is in general curved,   
and it should be necessarily curved 
if we want to describe the gravitational dynamics on the brane world. 
We adopt a curvilinear coordinate system $(x^0,x^1,x^2,x^3,y)$ 
such that the extra dimension coordinate $y$ vanishes on the brane.
Here we assume the whole spacetime is flat, for simplicity.
Then, the kink solution $\Phi_K(y)$  becomes an approximate solution for the domain wall. 
However, the equation of motion 
\begin{eqnarray}
\partial_M E G^{MN} \partial_N \Phi=-EU'(\Phi)  \label{em}
\end{eqnarray}
	with $E=\det E_{KM}$ $(K,M=0,\cdots,4)$ involves now 
	the vielbein $ E_{KM}$ and metric tensor $G_{MN}=E_{KM}{E^K}_N$
	for the curvilinear coordinate,
and the solution should be distorted from the kink function $\Phi_K(y)$ 
depending on the metric.
Here we assume that the extra dimension coordinate $y$ 
	is taken along the straight normal line 
	perpendicular to the brane at their crossing point. 
Then the bulk vielbein is given by
\begin{eqnarray}
\pmatrix{	E_{k\mu} &	E_{k4}\cr
		E_{4\mu} &	E_{44}	}
=\pmatrix{	e_{k\mu}-yb_{k\mu} 	&	0\cr
		0 			&	-1},
\end{eqnarray}
where $e_{k\mu}$ is the induced vielbein on the brane,  
and the $b_{\mu\nu}$ is the extrinsic curvature of the brane.
The vielbein and the extrinsic curvature should obey the Gauss-Codazzi equation
which is the embeddability condition of the brane.

We solve the equation of motion (\ref{em})
	by rewriting the field $\Phi$ into the form
\begin{eqnarray}
&&\Phi(x^{\mu},y)=\Phi_K(y)+\chi(x^{\mu},y)\Phi_k(y)',\cr
&&\chi(x^{\mu},y)=\chi_0(x^{\mu})+\chi_1(x^{\mu})y+\cdots.
\end{eqnarray}
Then the equation of motion reduces to recursion formulae for the coefficients $\chi_i(x^\mu)$.
We determine them one by one. 
The solution is given by
\begin{eqnarray}
&&\Phi=\Phi_{br}\equiv\Phi_K+\chi\Phi_k',\cr
&&\chi= by^2/2+(b_2+b^2)y^3/6 +\cdots.
\end{eqnarray}
	where $b=b_{\mu\nu}g^{\nu\mu}$,  
	$b_2=b_{\mu\nu} b_{\rho\sigma}g^{\nu\rho} g^{\sigma\mu}$
	and we put the arbitrary functions $\chi_0$ and $\chi_1$ 
	vanishing for simplicity.

We consider the quantum fluctuations around this classical solution $\Phi_{br}$.
For this purpose we expand our dynamical variable $\Phi$  
	in terms of the complete set $\{\eta_n\}$ 
	of small fluctuation modes of the kink solution $\Phi_K$. 
\begin{eqnarray}
\Phi=\Phi_{br}+\varphi_0\eta_0+\varphi_1\eta_1+\int\varphi_q\eta_qdq,
\label{Phi}
\end{eqnarray}
	where $\{\eta_n\}$ consists of 
\begin{eqnarray}
&& \eta_0=1/\cosh^2z,\ \ \ \ (z=my/\sqrt2)
\\&& \eta_1=\sinh z/\cosh^2z,
\\&& \eta_q=e^{iqz}(\tanh^2 z-1-q^2-3iq\tanh z.\ \ \ \ \ 
\end{eqnarray}
Among them, we omit the translation zero mode $\eta_0$, 
	because it only translates the position of the brane,     
	and should be considered with other configuration 
	of the brane vielbein $e_{k\mu}$ and the extrinsic curvature $b_{\mu\nu}$.
The dynamical degrees of freedom of the zero mode 
	are transferred to those of the brane vielbein and the extrinsic curvature
	which are constrained by the Gauss-Codazzi equation.

We substitute the expanded form (\ref{Phi}) of the field $\Phi$ 
	back into the original whole-spacetime action       
\begin{eqnarray}
\hskip-5mm
S=\!\!\int\!\! E\left(G^{MN}\partial_M\Phi\partial_N\Phi/2-U(\Phi)\right)d^5x,
\end{eqnarray}
	and rearrange the terms with respect to the dynamical variables 
	$e_{k\mu}$, $b_{\mu\nu}$, $\varphi_1$, and $\varphi_q$.     
Among the terms, those concerned with $e_{k\mu}$, $b_{\mu\nu}$, and $\varphi_1$ 
	are localized around the brane.
This means that the fields are trapped on the brane.
So we perform the $y$ integration for the trapped sector. 
\begin{eqnarray}
&&S=\int 
E\bigg[{1\over2}G^{MN}(\partial_M\varphi_1\eta_1) (\partial_M\varphi_1\eta_1)
\cr&&
-{m^4\over4\lambda }-{1\over2}m^2\varphi_1^2\eta_1^2
+{1\over4}\lambda(\Phi_K+\chi\Phi_K')^4
\cr&&
-{3\over2}\lambda(\Phi_K+\chi\Phi_K')^2\varphi_1^2\eta_1^2
\bigg]dyd^4x
\end{eqnarray}
They are integrations of the known functions such as 
\begin{eqnarray}
&&\int{\eta_1}^2dy={\sqrt2\over m}\int
\left({\sinh z\over\cosh^2z}\right)^2dz
={2\sqrt2\over3m},
\cr&&
\int y^4\Phi_K'^2\eta_1^2dy={2\sqrt2\over\lambda m}\int
{z^4\sinh^2 z\over\cosh^8z}dz
\cr&&
={2\sqrt2\over\lambda m}
\left(
-{2\over15}-{\pi^2\over45}+{\pi^4\over225}
\right),\ \rm etc.\nonumber
\end{eqnarray}
Thus we finally obtain the effective action 
\begin{eqnarray}
&&S=\int e \bigg[-{2\sqrt2m^3\over3\lambda}
+{\sqrt2m\over3\lambda}\left(1+{\pi^2\over3}\right)b_2
\cr&&
+{\sqrt2m\over\lambda }
\left(
-{4\over5}-{\pi^2\over6}+{7\pi^4\over1200}
\right)b^2
\cr&&
+{\sqrt2\over3m}\left(g^{\mu\nu}\partial_\mu\varphi_1\partial_\nu\varphi_1
-{3\over2}m^2{\varphi_1}^2
\right)
\cr&&
+{\sqrt2\over3m^3}\left(1+{\pi^2\over12 }\right)
\partial_\mu\varphi_1\partial_\nu\varphi_1
\cr&&
\times 
\{6b^\mu_{\ \lambda} b^{\lambda\nu}-4bb^{\mu\nu}
-(b_2-b^2)g^{\mu\nu}\}\cr&&
+{\sqrt2\over m}
\left(-{9\over10}+{\pi^2\over360}-{\pi^4\over300}\right)b^2{\varphi_1}^2
\cr&&
+{\sqrt2\over m}\left({5\over6}-{\pi^2\over360}\right)
b_{\mu\nu} b^{\mu\nu}{\varphi_1}^2
\bigg]d^4x
\end{eqnarray}
They consist of the cosmological term, the mass terms of the field $b_{\mu\nu}$,
	the kinetic and the mass terms of the field $\varphi_1$
and the interactions among them. 
The masses and coupling constants are definitely calculated.
The field $\varphi_1$ is a prototype of trapped field on the brane.      
If there exist bosonic or fermionic fields 
which coupled to our original field $\Phi$ in the whole spacetime,   
their low-lying modes are trapped around the brane,
and described by similar effective actions on the brane.

A problem is, however, no kinetic terms are induced       
for the vielbein $e_{k\mu}$ and the extrinsic curvature $b_{\mu\nu}$.
This is because we assumed that the whole space is flat for a technical simplicity.
If the whole space is curved, 
	the expression of the whole-space vielbein $E_{KM}$ involves 
	the derivatives of the brane vielbein $e_{k\mu}$ 
	and extrinsic curvature $b_{\mu\nu}$,
	which give rise to their kinetic terms. 
Among them the Einstein gravity action is induced for the brane metric $g_{\mu\nu}$.
This is nothing but the trapped graviton in its proper treatment.
An outstanding feature of the brane world picture is that
	the extrinsic curvature fields become dynamical within the brane
	under the constraint of the Gauss-Codazzi equation.
Anther effect which contributes to the kinetic terms     
are the quantum fluctuation of the trapped matters on the brane
which is calculated through the quantum loop diagrams     
with internal lines of these matter fields.\cite{Sakharov}$^-$\cite{Zee}
Interestingly, the kinetic terms due to the quantum effects          
can be induced even if the whole spacetime is flat.

So far we assumed a specific model of a 3 brane in a 4+1 dimensions.
The number of the extra dimension was 1.
If the number is grater than 1, 
the normal connections of the brane naturally take part in the play.
They are nothing but the gauge fields of the O(N) rotation group of the extraspace.
If we interpret them as the physical gauge fields like photon, gluon, and weak bosons,
	interesting phenomenological consequences may emerge,      
	since they are severely constrained by the Gauss-Codazzi-Ricci equations.

In conclusion, 
on the dynamically localized brane world:
(i) Low lying small fluctuation modes are trapped with proper field theoretical effective action. 
(ii) The gravitational field $e_{k\mu}$, 
	the extrinsic curvature $b_{\mu\nu}$, 
	and the gauge fields $A_\mu^{ab}$
	are induced on the brane.
(iii) The masses and coupling constants are definitely calculated. 
(iv) The kinetic terms of $e_{k\mu}$, $b_{\mu\nu}$, and $A_\mu^{ab}$ on the brane 
	are induced through quantum fluctuations of the trapped matters, 
	and they are induced even if the whole spacetime is flat.
(v) $e_{k\mu}$, $b_{\mu\nu}$, and $A_\mu^{ab}$ 
	should obey the Gauss-Codazzi-Ricci equations
	inaddition to their individual equations of motion.

\end{document}